\documentclass[aps,pre,amsmath,amssymb,twocolumn,groupedaddress]{revtex4-2}

\usepackage{hyperref}

\usepackage{graphicx}
\usepackage{dcolumn}
\usepackage{bm}

\newcommand{\anorm}[1]{\frac{#1 A}{\gamma b^2}}
\newcommand{\ainv}[1]{\frac{#1 \gamma b^2}{A}}

\begin{document}


\title{Persistent motion of a Brownian particle subject to repulsive feedback with time delay}
 
\author{Robin A. Kopp}
\email{r.kopp@tu-berlin.de}
\author{Sabine H. L. Klapp}
\email{sabine.klapp@tu-berlin.de}
\affiliation{%
Institut für Theoretische Physik, Technische Universität Berlin, Hardenbergstr. 36, D-10623 Berlin, Germany\\
}

\date{\today}

\begin{abstract}
Based on analytical and numerical calculations we study the dynamics of an overdamped colloidal particle moving in two dimensions under time-delayed, non-linear feedback control. Specifically, the particle is subject to a force derived from a repulsive Gaussian potential depending on
the difference between its instantaneous position, $\mathbf{r}(t)$, and its earlier position $\mathbf{r}(t-\tau)$, where $\tau$ is the delay time. Considering first the deterministic case, we provide analytical results for both, the case of small displacements and the dynamics at long times. In particular, at appropriate values of the feedback parameters, the particle approaches a steady state with a constant, non-zero velocity whose direction is constant as well. In the presence of noise, the direction of motion becomes randomized at long times, but the (numerically obtained) velocity autocorrelation still reveals some persistence of motion. Moreover, the mean-squared displacement (MSD) reveals a mixed regime at intermediate times with contributions of both, ballistic motion and diffusive translational motion, allowing us to extract an estimate for the effective propulsion velocity in presence of noise. We then analyze the data in terms of exact, known results for the MSD of active Brownian particles. The comparison indeed indicates a strong similarity between the dynamics of the particle under repulsive delayed feedback and active motion. This relation carries over to the behavior of the long-time diffusion coefficient $D_\mathrm{eff}$ which, similarly to active motion, is strongly enhanced compared to the free case. Finally we show that, for small delays, $D_\mathrm{eff}$ can be estimated analytically.
\end{abstract}

\maketitle
\newpage
\section{\label{sec:intro}Introduction}
Within the last years, feedback (closed-loop) control of colloidal systems,
that is, nano- to micron-sized particles in a thermally fluctuating bath of solvent particles, has become a focus of growing interest. Conceptually, feedback control implies that the dynamics of a system
is subject to a protocol depending on an internal variable, i.e., internal information
from the system. This concept is already widely used in other fields of physics (and related sciences) and on
many different length- and time scales \cite{bechhoefer_feedback_2005,scholl_control_2016}, examples ranging from the stabilization of exotic
quantum states and quantum computation over control of transport of (passive) colloids
to switching processes in neurosystems, applications in robotics and cars, and chaos control of
satellites.  In contrast, the use and design of feedback control in colloidal or, more generally, soft matter systems is a rather new development.
Recent experimental applications include 
control of DNA molecules by feedback-driven temperature fields \cite{braun_single_2015,Thalheim:17} or optical traps \cite{dieterich_control_2016}, magnetic feedback control of cells \cite{fisher_three-dimensional_2005},
electrophoretic feedback control of nanoparticles \cite{cohen_control_2005}, 
colloids in a electrokinetic feedback trap \cite{jun_high-precision_2014}, in optical line traps \cite{lopez_realization_2008}, and feedback cooling of nanoparticles \cite{gieseler_non-equilibrium_2015}. 

A new, exciting field of application is feedback control of active (self-propelled) colloids which move autonomously due to an intrinsic source of energy. Control of the translational or rotational motion
of active colloids can be realized, e.g., by photon nudging and by adaptive light fields \cite{qian_harnessing_2013,bregulla_stochastic_2014,selmke_theory_2018-1,selmke_theory_2018,franzl_active_2020, haeufle_external_2016,mijalkov_engineering_2016,fernandez-rodriguez_feedback-controlled_2020,bauerle_formation_2020-1,loffler_behavior-dependent_2021}. Interest in this topic is triggered by the immense body of work on active particles showing intriguing collective behaviors such as swarming and clustering. 
Feedback control of such systems offers a way of modelling living matter involving not only autonomous motion, but also communication, sensing and thereby new types of self-organization. 
In addition to these efforts, feedback control of colloids is interesting, on
a fundamental level, to study its interplay with thermo-
dynamics and information exchange in small stochastic
systems \cite{toyabe_experimental_2010,blickle_realization_2012,koski_experimental_2014,jun_high-precision_2014,loos_irreversibility_2020,cao_thermodynamics_2009} on the basis of stochastic thermodynamics \cite{ seifert_stochastic_2012}.

In any realistic setup of feedback control (and similarly in biological systems), there is some time lag (or {\em time delay}) between the reception of information (e.g., via a camera) and the actual control. Similarly, biological living systems with feedback mechanisms often exhibit some degree of sensorial delay \cite{mijalkov_engineering_2016}, communication delay \cite{hindes_hybrid_2016}, or, more generally, memory effects due to viscoelastic environments \cite{clark_self-generated_2022}. Thus, the idea of {\em instantaneous} feedback is often an idealization. Moreover, there is now increasing awareness that this time delay is not {\em per se} an annoyance, but can rather be an important ingredient to observe and stabilize certain dynamical behavior. The constructive role of time delay has already been shown in other contexts \cite{bechhoefer_feedback_2005, scholl_control_2016,scholl_eckehard_handbook_2007} including chaos control \cite{pyragas_continuous_1992, pyragas_control_1995}, \cite{tsui_control_2000}, laser systems \cite{bielawski_controlling_1994, marino_pseudo-spatial_2017}, chemical oscillatory systems \cite{schneider_continuous_1993} and reaction networks \cite{thanh_efficient_2017}. 
Recently is has been shown that time delay has also important consequences for the thermodynamics of feedback-controlled systems \cite{debiossac_thermodynamics_2020, loos_irreversibility_2020, loos_heat_2019}. 

In colloidal systems,
time-delayed feedback has already been shown to create intriguing effects. Examples in the field of transport of (one-dimensional) passive colloidal systems include current reversals \cite{lichtner_feedback-controlled_2010} and enhancements  \cite{lopez_realization_2008,loos_delay-induced_2014,gernert_enhancement_2015} in ratchet and tilted washboard systems, but also band formation in two-dimensional interacting colloidal systems with collective time-delayed feedback \cite{tarama_traveling_2019}. Moreover, recent studies of (time-delayed) feedback control in {\em active} systems \cite{selmke_theory_2018-1, selmke_theory_2018, franzl_active_2020} have revealed new effects such as delay-induced clustering and swarming \cite{mijalkov_engineering_2016,holubec_finite-size_2021-1,sprenger_active_2022,wang_spontaneous_2022}.
Despite the increasing awareness of the relevance of delay and feedback effects in colloidal systems, theoretical studies of specific model systems are still rare. This is mainly due to the non-Markovianity of the underlying stochastic equations of motion, making explicit calculations challenging. Indeed, exact results only exist for {\em linear} models \cite{kuchler_langevins_1992-1, guillouzic_small_1999-1,frank_stationary_2001-2,frank_fokker-planck_2003-1}. However, in real colloidal systems, the interactions and forces involved are typically nonlinear. 
Motivated by these developments, we here present a theoretical study of a particular nonlinear model of a colloidal particle under time-delayed feedback. Specifically, we consider the two-dimensional translational motion of a single, spherical colloid subject to a repulsive force generated by a Gaussian potential centered around the particle position at an earlier time $t-\tau$. Thus, the position acts as a stochastic variable on which the control is performed. 
Our model differs in several aspects from earlier ones (see, e.g. \cite{loos_delay-induced_2014,gernert_enhancement_2015}). We here consider {\em repulsive} feedback, rather than the often studied case of trapping a particle by an attractive potential (stemming, e.g., from an optical tweezer). We note that similar nudging mechanisms have already been used in the context of the ``photo-nudging" method for Janus particles \cite{qian_harnessing_2013,bregulla_stochastic_2014,selmke_theory_2018-1,selmke_theory_2018,franzl_active_2020,franzl_fully_2021}. A second difference is that the force is nonlinear in the particle position and finite in range. Thus, although our model is not directly designed to describe a specific experiment and is, in this sense, hypothetical, it allows us to focus exclusively on several important features of a control scheme for Brownian particles: time-delay, repulsion, and nonlinearity.

Explicit results are obtained based on analytical considerations in some limiting cases, as well as on numerical solution of the corresponding (overdamped) non-Markovian Langevin equation. The first main result of our study is that for appropriate values of delay time and strength of repulsion, and in the absence of noise, the particle develops a stationary state characterized by a constant velocity vector. Second, thermal fluctuations lead to a randomization of the direction of motion, but the particle still possesses a persistence of motion and a constant magnitude of speed (after noise-averaging). This is reminiscent of the stochastic behavior of self-propelled particles, and indeed we identify close similarities. In particular, we discuss relations to the prominent model of active Brownian particles (ABP) \cite{lowen_inertial_2020}.
	For this model, which is widely used also to explain experimental data (see, e.g. \cite{buttinoni_active_2012-1, bechinger_active_2016}), there exists a closed analytical expression for the mean-squared displacement which we utilize to fit and interpret our numerical data. Further, the ABP model is one of the best studied systems concerning the collective behavior induced by activity \cite{zottl_emergent_2016-1,bechinger_active_2016}. Thus, establishing a link to the ABP model will provide us with a basis for future investigation of the collective behavior of our model. By investigating, in the present study, the single-particle relations between the different models we make a first step to establishing a link between feed-back controlled and active matter.

The rest of the paper is organized as follows: In the following section \ref{sec:model} we introduce our model and the corresponding overdamped Langevin equation governing the dynamics of the system. Subsequently we study, first, the deterministic limit in Sec. \ref{sec:deterministic}, using a combination of analytical and numerical methods. We then present our results for the full, nonlinear stochastic system, based on Brownian dynamics simulations, in Sec. \ref{sec:noisy}. 
Our conclusions are summarized in Sec. \ref{sec:conclusions}.
\section{\label{sec:model}Model}
	We consider the two-dimensional motion of a Brownian particle in the $x$-$y$ plane. In addition to thermal fluctuations due to a coupled heat bath at temperature $T$, the particle is subject to a time-delayed feedback force
	$\mathbf{F}$ depending on both, its instantaneous position $\mathbf{r}(t)= (x(t),y(t))^T$ and its position at an earlier time, $\mathbf{r}(t-\tau)$, where $\tau$ is the (discrete) delay time. The particle's motion at times $t>0$ is governed by the overdamped Langevin equation
	\begin{equation}
		\gamma \frac{d\mathbf{r}}{dt} = \mathbf{F}\left(\mathbf{r}(t),\mathbf{r}(t-\tau)\right)+\bm{\xi}(t),
		\label{eq:delaylangevin}
	\end{equation}
	where $\gamma$ is the friction coefficient, and the position at earlier times $t\in[-\tau,0]$ is determined by the history function $\bm{\Phi}(t)$. 
	Furthermore, $\bm{\xi}$ represents a two-dimensional, Gaussian white noise with zero mean and correlation function
	$\langle \xi_\alpha (t) \xi_\beta(t^\prime) \rangle = 2 \gamma k_BT\delta_{\alpha \beta} \delta(t-t^\prime)$
	where $\alpha$, $\beta$ are the cartesian components of the noise vector $\bm{\xi}$, and $k_B T$ (with $k_B$ being the Boltzmann constant) is the thermal energy. The diffusion constant of the free particle
	motion ($\mathbf{F}=\mathbf{0}$) follows from the Stokes-Einstein relation $D=k_B T/\gamma$.
	
\begin{figure}
\centering
	\includegraphics[width = 0.45\textwidth]{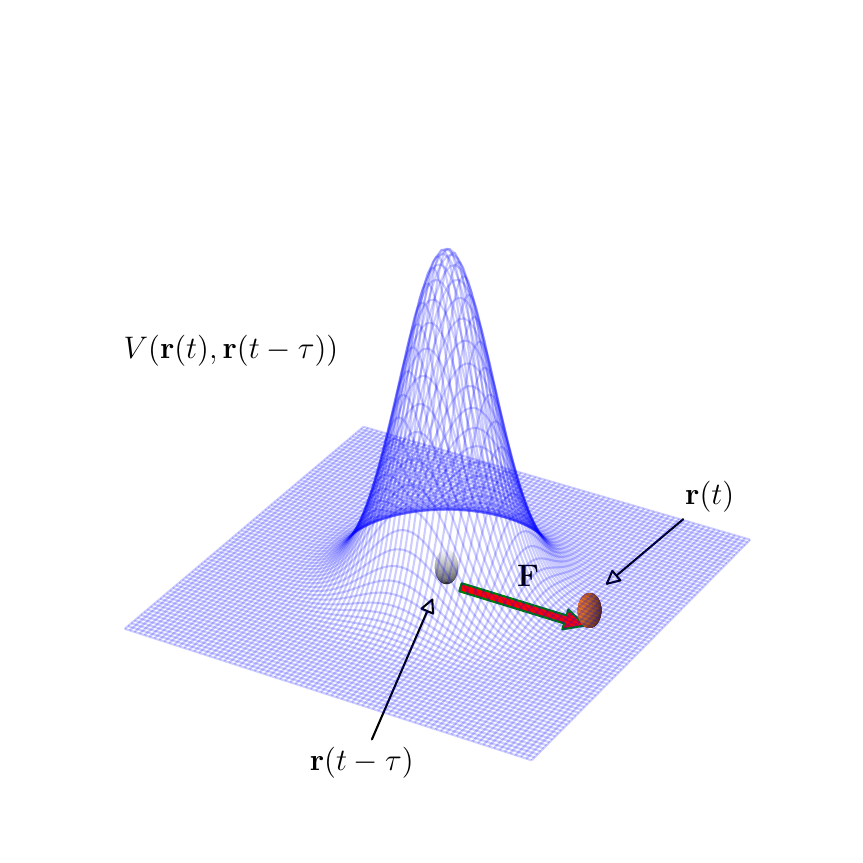}
	\caption{Schematic of the repulsive Gaussian delayed feedback potential $V$. At each time $t$, the Gaussian is centered around the earlier position $\mathbf{r}(t-\tau)$, resulting in a ``pushing" force (red arrow) directed along the difference vector $\mathbf{r}(t)-\mathbf{r}(t-\tau)$.\label{fig:potential}}
\end{figure}

Within our model, the feedback force $\mathbf{F}$ is derived from a Gaussian feedback potential involving the displacement $\mathbf{r}(t) - \mathbf{r}(t-\tau)$, that is,
	\begin{equation} 
		V(\mathbf{r}(t),\mathbf{r}(t-\tau)) = A \exp\left(-\frac{\left(\mathbf{r}(t) - \mathbf{r}(t-\tau)\right)^2}{2b^2}\right),
		\label{eq:gauss}
	\end{equation}
yielding
	\begin{eqnarray}
		\mathbf{F} & = &-\nabla_{\mathbf{r}} V(\mathbf{r}(t),\mathbf{r}(t-\tau))\nonumber\\
		& = & \frac{A}{b^2}(\mathbf{r}(t) - \mathbf{r}(t-\tau)) \exp\left(-\frac{\left(\mathbf{r}(t) - \mathbf{r}(t-\tau)\right)^2}{2b^2}\right).
		\label{eq:fbforce}
	\end{eqnarray}
We choose the feedback strength $A > 0$, representing {\em repulsive} feedback whose range is determined by the width of the Gaussian, $b$ ($b>0$). The width of the feedback potential serves as the length scale in our system. An illustration of the ``bump"-like potential is given in Fig.~\ref{fig:potential}.
Physically, 
the potential (\ref{eq:gauss}) can be interpreted as a source of a ``nudge" following the particle with some delay.  Such a potential could, in principle, be created by optical forces  \cite{bauerle_self-organization_2018,leyman_tuning_2018,lavergne_group_2019,franzl_active_2020,franzl_fully_2021,bauerle_formation_2020-1}.
We note that, different to the often used harmonic potentials leading to linear feedback forces \cite{loos_medium_2021}, the potential (\ref{eq:gauss}) and the force (\ref{eq:fbforce}) decay to zero when the displacement within one delay time, $\mathbf{r}(t)-\mathbf{r}(t-\tau)$, increases to large values. The finite range is indeed advantageous when using Eq.~(\ref{eq:gauss}) in a many-particle system with periodic boundary conditions. 
We also note, from the perspective of control theory, the feedback force has ``Pyragas" form \cite{pyragas_continuous_1992, pyragas_control_1995} 
since it only depends on the difference between $\mathbf{r}(t)$ and $\mathbf{r}(t-\tau)$. Thus, the force vanishes trivially if $\tau=0$; in this case, our model reduces to that of a free Brownian particle.
Note, however, that the feedback force also vanishes when the particle is at rest, that is,  $\mathbf{r}(t-\tau)=\mathbf{r}(t)$, or when it moves in a cyclic fashion with period $\tau$.

Inserting the expression for the force (\ref{eq:fbforce}) into Eq.~(\ref{eq:delaylangevin}), the complete equation of motion reads
		\begin{eqnarray}
			\gamma\frac{d\mathbf{r}}{dt}& =& \frac{A}{b^2}(\mathbf{r}(t) - \mathbf{r}(t-\tau)) \exp\left(-\frac{\left(\mathbf{r}(t) - \mathbf{r}(t-\tau)\right)^2}{2b^2}\right) \nonumber\\
			&  &+ \bm{\xi}(t).
			\label{eq:sdde}	  
		\end{eqnarray}
Due to the presence of time delay, the particle’s dynamics defined by Eq. (4) is strongly non-Markovian (where ``strongly" refers to the fact that the kernel emerging when we write the right side as a convolution over past times is 
a delta peak located at a finite time $\tau$ \cite{loos_fokkerplanck_2019}). As a consequence, several tools well established for Markovian Brownian systems (such as the link between Langevin and Fokker Planck descriptions) do not straightforwardly apply.
More mathematically spoken, Eq.~(\ref{eq:sdde}) represents a stochastic delay differential equation (SDDE) which is, furthermore, non-linear due to the Gaussian shape of the underlying potential. 
Even in the absence of noise, 
the resulting differential equation (DDE) formally becomes infinite-dimensional due to the presence of the continuous history function $\bm\Phi(t)$. 
In recent years, a number of results have been obtained for deterministic DDEs \cite{driver_ordinary_1977,erneux_applied_2009-1,atay_complex_2010}, 
as well as for {\em linear} SDDEs see, e.g., \cite{kuchler_langevins_1992-1,guillouzic_small_1999-1,frank_stationary_2001-2,frank_fokker-planck_2003-1}. However, here we are dealing with a nonlinear SDDE for which, to our knowledge, no {\em full} analytical solution exists.
Still, we can analytically consider some limiting cases, which we will discuss in the subsequent Sections~\ref{sec:deterministic} and \ref{sec:noisy}. In addition, we study the particle dynamics given by Eq.~(\ref{eq:sdde}) and its deterministic limit numerically using Brownian Dynamics (BD) simulations. Some technical details of the numerical calculations are given in Appendix~\ref{sec:technical}.
%
%

\section{\label{sec:deterministic}Deterministic limit}
We start by considering the deterministic limit of Eq.~(\ref{eq:sdde}) defined by $\bm{\xi}=\mathbf{0}$, which may be realized by setting the temperature $T$ to zero. In this way we can explore the role of delay alone, thereby providing a useful starting point for the later investigation of the noisy case.
\subsection{\label{sec:linear}Linear behavior}
\begin{figure}
 \centering
\includegraphics[width=.45\textwidth]{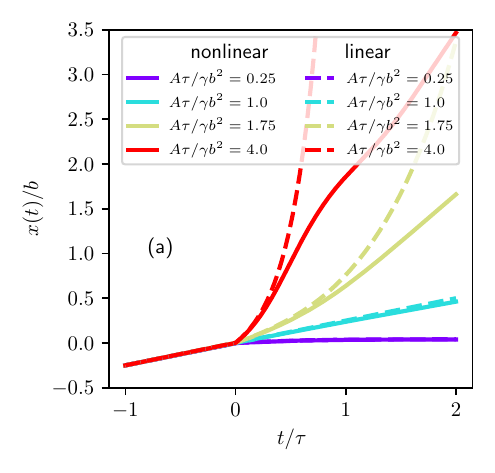}
\includegraphics[width=.45\textwidth]{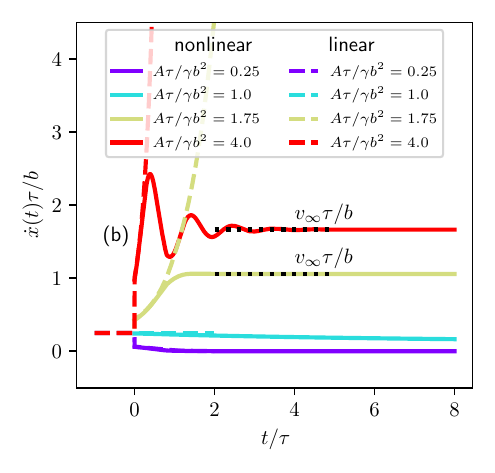}
\caption{Numerical results for the time dependence of the position (a) and velocity (b) of the particle subject to nonlinear, Gaussian feedback in the deterministic limit (for a history with constant (one-dimensional) velocity 
$\mathbf{v}_0=v_0\hat{\mathbf{x}}$, i.e., $\bm{\Phi}=\mathbf{v}_0t$). Included are the analytical results for the linearized case (small displacements [see Eq.~(\ref{eq:linear})]).
The curves are labelled according to the value of the dimensionless parameter $A\tau/\gamma b^2$.\label{fig:deterministic}}
\end{figure}

Some first insights can already be obtained by investigating the behavior at small displacements $|\mathbf{r}(t)-\mathbf{r}(t-\tau)|\ll \sqrt{2}b$. In this case, we can expand the force (\ref{eq:fbforce}) up to first order in the displacement, yielding the linear equation
	\begin{equation}
		\gamma \dot{\mathbf{r}} = \frac{A}{b^2}\left(\mathbf{r}(t)-\mathbf{r}(t-\tau)\right).
		\label{eq:linear}
	\end{equation}
In deriving Eq.~(\ref{eq:linear})  we have used that the Hessian matrix $\mathbf{H}(\mathbf{r})$ with elements $H_{\alpha\beta}(\mathbf{r})=\partial^2V/\partial x_\alpha\partial x_\beta$ (with $V(\mathbf{r})$ being the feedback potential) 
is diagonal at $\mathbf{r}=0$, and $H_{\alpha\alpha}(\mathbf{0})=-A/b^2$. As a consequence of the diagonality, each component $x_\alpha$ of $\mathbf{r}$ can be considered separately. The resulting scalar, linear DDEs are indeed well studied.
For example, it is well known \cite{liu_stability_1984,hovel_control_2005, asl_analysis_2003} that the (fixed) point $\mathbf{r}=\mathbf{r}_0$ with $x_\alpha=const$ is marginally stable in the sense that a perturbation $\propto \exp[\mu t]$ (with $\mu$ being a complex number) remains constant, that is, $\mu=0$. Physically, this expresses the metastability of a particle on a ``parabolic mountain" \cite{loos_medium_2021}. Moreover,
explicit solutions of the linear equation~(\ref{eq:linear}) can be constructed in a piecewise manner by using the {\em method of steps} \cite{driver_ordinary_1977,erneux_applied_2009-1}. 
For a given history function
$\bm{\Phi}(t)$ with components ${\Phi}_\alpha(t)$ (defined in the interval $t\in[-\tau,0]$), the solution in the first interval $t\in [0,\tau]$ follows as
	\begin{eqnarray}
		x_\alpha & = & \exp\left(\frac{At}{\gamma b^2}\right)\left ( \Phi_\alpha (0)\right.\nonumber\\
		&  &\left.-\frac{A}{\gamma b^2}\int_0^tdt'\exp\left(\frac{A\left( - t'\right)}{\gamma b^2}\right) \Phi_\alpha(t'-\tau)\right )
		\label{eq:solution_linear}
	\end{eqnarray}
which shows directly the dependency on the history. In particular, if the particle was at rest at some fixed position ($\bm{\Phi}=\mathbf{r}_0$), it stays where it is (as expected from the marginal stability) for all values of the prefactor 
$A/b^2$. This then also holds for the later time intervals $[n\tau,(n+1)\tau]$ (with $n=1,2,\ldots$). A more interesting situation for the present study occurs if the particle had moved with a constant velocity $\mathbf{v}_0$, that is,
$\bm{\Phi}=\mathbf{v}_0t$, $t\in[-\tau,0]$. The corresponding explicit solution for $t>0$ up to time $2\tau$ is given in the Appendix~\ref{sec:appendix_solution}.
It turns out that the behavior of $x_\alpha(t)$ (and the corresponding velocity) crucially depends on the value of the dimensionless parameter $A\tau/\gamma b^2$. If this parameter is smaller than one, the particle just approaches a constant position 
and the velocity dies off to zero. In contrast, if $A\tau/\gamma b^2>1$, the position and the corresponding velocity continue to increase unboundly. Finally, at the ``threshold" value $A\tau/\gamma b^2=1$, the behavior just remains identical to that given by the history, that is, the particle continues to move with the velocity $\mathbf{v}_0$. An illustration of these behaviors  is given in Fig.~\ref{fig:deterministic} (dashed lines), where the history is characterized by a one-dimensional velocity (generating one-dimensional motion), $\mathbf{v}_0=v_0\hat{\mathbf{x}}$, $t\in[-\tau,0]$.
\subsection{\label{sec:longtime}Nonlinear behavior and steady state}
We now consider the full, nonlinear (deterministic) system. The corresponding DDE (\ref{eq:sdde}) with $\bm{\xi}=\mathbf{0}$ cannot be solved any more by the method of steps, 
such that we utilize a numerical solution to obtain the full trajectory $\mathbf{r}(t)$
in dependence of the history $\bm{\Phi}(t)$. Examples are shown in Fig.~\ref{fig:deterministic} (solid lines) where we focus (again) on an initial velocity along the $x$-axis. At small times $t/\tau>0$, the solution agrees with that obtained for the (linear) case describing small displacements, as it should do. The further time dependence of the nonlinear system again crucially depends on the dimensionless parameter $A\tau/\gamma b^2$. In particular, for $A\tau/\gamma b^2\leq 1$ the nonlinear system comes to rest in the sense that the position settles and the velocity vanishes. Note that this is consistent with the linear case for 
$A\tau/\gamma b^2< 1$, but not for the ``threshold" value $A\tau/\gamma b^2=1$, where the linear system is fully governed by the history.
Moreover, a crucial difference between the systems emerges when $A\tau/\gamma b^2>1$.
Here, the nonlinear system governed by the full, Gaussian feedback develops a stationary state characterized by a {\em constant} velocity $\mathbf{v}_{\infty}$. This velocity can be determined analytically as follows.

Let us assume that such a state develops. We then have $\dot{\mathbf{r}}=\mathbf{v}_\infty$ and $\mathbf{r}(t) - \mathbf{r}(t-\tau) = \mathbf{v}_\infty \tau$. Inserting this into the deterministic version
of Eq.~(\ref{eq:sdde}) we obtain an implicit equation for the long-time velocity, 	  
\begin{equation}
\mathbf{v}_\infty= \frac{A\tau}{\gamma b^2} \mathbf{v}_\infty  \exp\left(-\frac{\left(\mathbf{v}_\infty\tau\right)^2}{2b^2}\right),
\label{eq:vinfty}
\end{equation}
yielding
\begin{equation}
|\mathbf{v}_\infty| = \pm\frac{\sqrt{2}b}{\tau}\sqrt{-\ln\left(\frac{\gamma b^2}{A\tau}\right)}.
\label{eq:const_vel}
\end{equation}
Equation~(\ref{eq:const_vel}) provides an explicit expression for the magnitude of the long-time velocity in terms of the parameters of the feedback potential.
Clearly, a real (and, thus, physically meaningful) value of $|\mathbf{v}_\infty|$ can only occur for $A\tau/ \gamma b^2>1$, providing a restriction for the minimal strength of the feedback potential, $A$, and/or the minimal delay time, $\tau$ (for given $b$ and $\gamma$). This result for the ``threshold" value
$A\tau/ \gamma b^2=1$ is consistent
with our numerical data shown in Fig.~\ref{fig:deterministic}. From Fig.~\ref{fig:deterministic} we also observe that for large values of the parameter (such as the case $A\tau/\gamma b^2=4.0$), the velocity approaches its stationary value only after a transient regime characterized by oscillations. 

	These oscillations can be understood as follows. After the onset of (strong) feedback, the resulting velocity $\mathbf{v}(t)$ is initially so large that the magnitude of the displacement within one delay time, $|\mathbf{r}(t)-\mathbf{r}(t-\tau)|$, first increases with $t$. This increase yields, in turn, an increase of the feedback force $\mathbf{F}$. Note that the latter is directly visible from Fig.~2(b) when we recall that, in the deterministic case, force and velocity are just proportional to one another. Notably, the increase of the force with $t$ comes to an end as soon as
	the displacement has reached the value corresponding to the {\em maximum} of the feedback force (related to the strongest descent of the feedback potential). 
	This occurs when $|\mathbf{r}(t)-\mathbf{r}(t-\tau)|=b$, with the corresponding maximal force being given by $|\bm{F}|_{\mathrm{max}}=(A/b)\exp(-1/2)$. The corresponding maximal value of the velocity 
	then follows (in dimensionless form) as $v_{\mathrm{max}}\tau/b=(A\tau/\gamma b^2)\exp(-1/2)\approx 0.6 (A\tau/\gamma b^2)$, consistent with the data (for, e.g., $A\tau/\gamma b^2=4.0$) in Fig.~2(b).
	Subsequently, the force and thus, the velocity decreases, yielding a slower increase of the actual position relative to the delayed one, and thereby, a decrease of the displacement $|\mathbf{r}(t)-\mathbf{r}(t-\tau)|$. However, at some point the situation reverses. This happens when the history ``kicks in", that is, 
	the before-mentioned decrease of the force (or velocity) becomes reflected by a slower change of the quantity $\mathbf{r}(t-\tau)$. Then the relative displacement $|\mathbf{r}(t)-\mathbf{r}(t-\tau)|$ increases again, yielding the second rise of the force (or velocity) with $t$ and so on.
	To summarize, the oscillations are essentially a consequence of the fact that the feedback force has a maximum at finite displacements (reflecting the nonlinearity) and is history-dependent.
	Clearly, oscillations becomes visible only when the maximum velocity (related to the maximum force) is sufficiently {\em different} from the long-time velocity, i.e., $v_{\mathrm{max}}\tau/b\neq v_\infty\tau/b$.

\begin{figure}
	\centering
	\includegraphics[width = 0.45\textwidth]{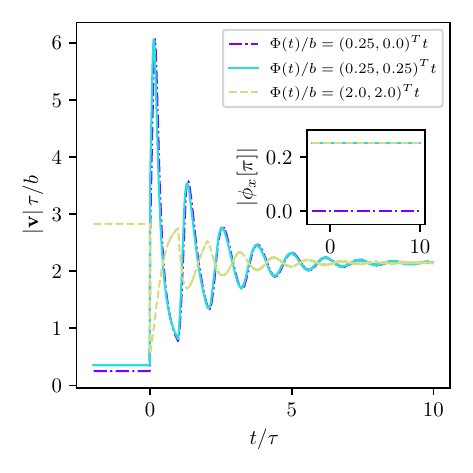}
	\caption{Numerical results for the time dependence of the particle velocity in the deterministic limit for three different choices of the constant vector $\mathbf{v}_0$ governing the history
	($\bm{\Phi}=\mathbf{v}_0t$). Main part: magnitude of the velocity, inset: polar angle (in units of $\pi$) between the velocity vector and the $x$-axis.\label{fig:history}}
\end{figure}

Importantly, the {\em magnitude} of the long-time velocity, $|\mathbf{v}_\infty|$, is essentially {\em independent} of the history $\bm{\Phi}(t)$ ($t\in[-\tau,0]$). This is shown in Fig.~\ref{fig:history} where we present numerical data for the velocity of the deterministic system for different choices of the constant velocity governing the history function $\bm{\Phi}=\mathbf{v}_0t$. Further, as indicated by extensive test calculations, this independence persists for history functions with more involved time dependence.
An exception is the case $\bm{\Phi}(t)=const$ (particle stays at rest) which leads to $|\mathbf{v}_\infty|=0$. 
In contrast to the magnitude, the direction of $\mathbf{v}_\infty$ is determined by history. This is illustrated by the inset of Fig.~\ref{fig:history} where we present data for the polar angle $\phi_x(t)$ enclosed by the instantaneous velocity and the $x$-axis. Besides the impact of the history value of the angle, the data also imply that the predescribed direction survives in the long-time limit. 

Given the rather robust appearance of a steady state with non-zero, constant velocity, it is interesting to investigate the stability of this state using the toolbox of linear stability analysis. To this end we consider the behavior of
a small deviation $\mathbf{s}(t):= \mathbf{r}(t)-\mathbf{v}_\infty t$, with $\dot{\mathbf{s}}=\dot{\mathbf{r}}-\mathbf{v}_\infty$. The displacement occurring in the feedback force then becomes
$\mathbf{r}(t)-\mathbf{r}(t-\tau)=\mathbf{s}(t)-\mathbf{s}(t-\tau)+\mathbf{v}_\infty \tau$. We now expand the nonlinear force around the constant-velocity state assuming that
both, $\mathbf{s}(t)$ and $\mathbf{s}(t-\tau)$ are small. Keeping only terms linear in $\mathbf{s}$ we obtain
\begin{eqnarray}
\gamma\left(\dot{\mathbf{s}}+\mathbf{v}_\infty \right) & = &\gamma\mathbf{v}_\infty\nonumber\\
& & - \mathbf{H}(\mathbf{v}_\infty\tau)\left(\mathbf{s}(t)-\mathbf{s}(t-\tau)\right).
\label{eq:sdot}
\end{eqnarray}
In the first term on the right side of Eq.~(\ref{eq:sdot}), we have used Eq.~(\ref{eq:vinfty}) to replace the force at zero perturbation. 
Clearly, this term cancels with the left side, yielding
\begin{eqnarray}
	\gamma\dot{\mathbf{s}} = - \mathbf{H}(\mathbf{v}_\infty\tau)\left(\mathbf{s}(t)-\mathbf{s}(t-\tau)\right),
	\label{eq:sdot2}
\end{eqnarray}
where the Hessian $\mathbf{H}$ contains the second-order derivatives of the feedback potential [see text below Eq.~(\ref{eq:linear})], evaluated in the stationary state.
The elements of the matrix $\mathbf{H}(\mathbf{v}_\infty \tau)$ thus read:
	\begin{eqnarray}
		H_{xx}(\mathbf{v}_{\infty}\tau)
		& = & -
		\frac{A}{b^2}\exp\left(-\frac{\left( \mathbf{v}_{\infty}\tau\right)^2}{2b^2}\right) \left(1- \frac{\left(v_{x,\infty} \tau\right)^2}{b^2}\right),\\
		H_{yy}(\mathbf{v}_{\infty}\tau) 
		& = & -
		\frac{A}{b^2}\exp\left(-\frac{\left( \mathbf{v}_{\infty}\tau\right)^2}{2b^2}\right) \left(1- \frac{\left(v_{y,\infty} \tau\right)^2}{b^2}\right),\nonumber\\
		H_{xy}(\mathbf{v}_{\infty}\tau)\nonumber &=& H_{yx}(\mathbf{v}_{\infty}\tau) \nonumber\\ 
		& = &
		\frac{A}{b^2}\exp\left(-\frac{\left( \mathbf{v}_{\infty}\tau\right)^2}{2b^2}\right) \left( \left(v_{x,\infty} \tau\right)\frac{\left(v_{y,\infty} \tau\right)}{b^2}\right).\nonumber
		\label{eq:Hessian} 
	\end{eqnarray}
	Clearly, this matrix is not diagonal; however, the $x$- and $y$ components decouple if $\mathbf{v}_\infty$ points along one of the axes of the coordinate system.
Without loss of generality, we assume that $\mathbf{v}_{\infty}=v_{x,\infty}\hat{\mathbf{x}}$. We then obtain
\begin{equation}
\dot{s}_x(t)=K\left(s_x(t) - s_x(t-\tau)\right)
\end{equation}
where
\begin{eqnarray}
K&=&-\gamma^{-1}H_{xx}(\mathbf{v}_{\infty}\tau)\nonumber\\
& = &
\frac{A}{\gamma b^2}\exp\left(-\frac{\left( v_{x,\infty}\tau\right)^2}{2b^2}\right) \left(1- \frac{\left(v_{x,\infty} \tau\right)^2}{b^2}\right).
\label{eq:K}
\end{eqnarray}
The expression in brackets can be simplified using Eq.~(\ref{eq:const_vel}), yielding
\begin{equation}
K=\frac{A}{\gamma b^2}\exp\left(-\frac{\left( v_{x,\infty}\tau\right)^2}{2b^2}\right) \left(1+2\ln\frac{\gamma b^2}{A\tau}\right).
\label{eq:K2}
\end{equation}
We note that the value of the logarithm must be negative (to ensure a real, non-zero value of the velocity $v_\infty$ in the long-time limit [see Eq. (\ref{eq:const_vel})]).
To perform the linear stability analysis, following \cite{hovel_control_2005, asl_analysis_2003}, we use the ansatz $s(t) = C\exp(\lambda t)$, where in general $\lambda \in \mathbb{C}$. 
This leads to the characteristic equation
	\begin{equation}
		\lambda = K(1-\exp(-\lambda \tau)).
	\end{equation}
	The solution of this transcendental equation, i.e. the eigenvalues $\lambda$ of the characteristic function, can be expressed via the Lambert function W \cite{asl_analysis_2003}	
	\begin{equation}
		\lambda = \frac{1}{\tau}W(-K\tau\exp(-K\tau)) + K.
	\end{equation}
	By applying an identity of the Lambert function W, namely $W(x\exp(x)) = x$ for $x\in\mathbb{R}$, we find that in our case the largest eigenvalue is always zero. Therefore, the {\em linear}
	stability analysis predicts the long-time state to be marginally stable. In Appendix~\ref{sec:perturbations} we present some numerical data illustrating the impact of a perturbation within the steady state. 
	For all cases considered, it turns out that the perturbations do not destroy the magnitude of the long-time velocity $\mathbf{v}_{\infty}$, but can change its direction.
	%
%
\section{\label{sec:noisy}Stochastic motion}
So far we have focused on the deterministic limit of Eq.~(\ref{eq:sdde}). We now turn to the major target of this paper, that is, the behavior of a {\em Brownian} particle subject to the repulsive, non-linear, time-delayed feedback force plus noise
in two dimensions.
Clearly, this situation calls for numerical investigation (for technical details, see Appendix~\ref{sec:technical}). Henceforth, we always consider history functions defined by the trajectory of a free Brownian particle. 
The dimensionless diffusion constant (noise strength) is set to $D\tau_B/b^2=1$, where the ``Brownian" time $\tau_B$, which we here define as the time a Brownian particle needs to diffuse over the length $b$, reads $\tau_B = b^2/D$.
\subsection{\label{sec:velocity}Trajectories and velocity correlations}
\begin{figure}
	\centering
	\includegraphics[width = 0.45\textwidth]{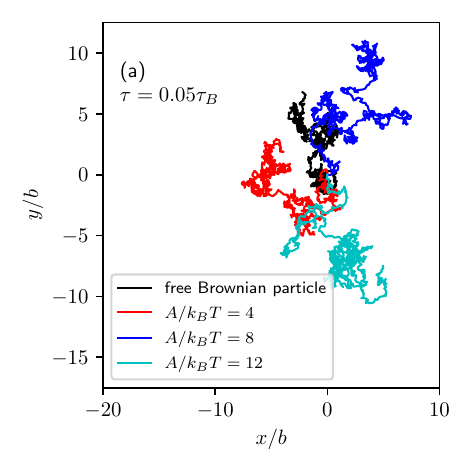}
	\includegraphics[width = 0.45\textwidth]{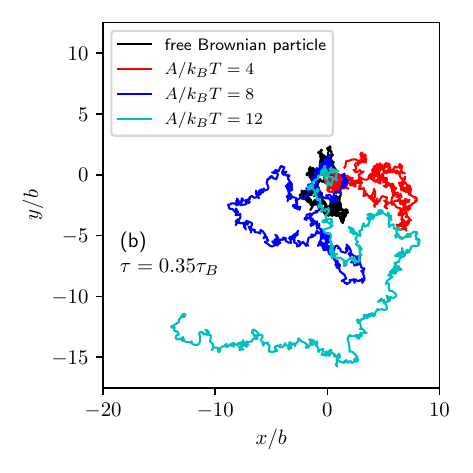}	
	\caption{Exemplary trajectories of Brownian particles subject to time-delayed feedback at different values of $A/k_B T$ in the time interval $[0, 10\tau_B]$. The case $A/k_BT=0$ corresponds to a free Brownian particle. The delay times are $\tau = 0.05\tau_B$ (a) and $\tau = 0.35\tau_B$ (b) respectively. In each case, the history $\mathbf{\Phi}(t)$, $t\in[-\tau,0]$, is given by a (randomly selected) trajectory of a free Brownian particle.\label{fig:trajectory}
	}
\end{figure}

In Fig.~\ref{fig:trajectory}, we present exemplary trajectories of the particle for two values of the delay time $\tau$ and various values of the dimensionless repulsion parameter $A/k_BT$ measuring the feedback strength relative to the thermal energy. Also shown is the corresponding ``free" Brownian particle ($A=0$). As seen for both values of $\tau$, a main effect of increasing $A/k_BT$ from zero is that the particle ``spreads" out more and more. Moreover, in the case $\tau=0.35\tau_B$ one observes long stretches where the particle travels along a preferred direction (see Fig.~\ref{fig:trajectory}~(b)).
In other words, the motion displays some {\em persistence} in it's direction of motion over a certain period of time. 

To better understand the impact of $\tau$ we recall that, in the deterministic case, the particle can develop a stationary state characterized by a non-vanishing 
magnitude and a constant (polar) angle of the velocity $\mathbf{v}_\infty$ provided that the dimensionless parameter $A\tau/\gamma b^2$ exceeds the value of $1$ (see Sec.~\ref{sec:deterministic}).
For the smaller delay time $\tau = 0.05\tau_B$, the maximal value of $A/k_B T=12$ considered in Fig.~\ref{fig:trajectory}~(a) 
corresponds to $A\tau/\gamma b^2=(A/k_BT)(\tau/\tau_B)(D\tau_B/b^2)=0.6$. Thus, the corresponding deterministic system would not develop any long-time velocity vector. In contrast, for $\tau = 0.35\tau_B$, the parameter $A\tau/\gamma b^2$ varies between $1.4$ and $4.2$, leading to a deterministic long-time motion with given velocity magnitude and direction.
Clearly, in the presence of noise, one would expect any preferred {\em direction} of motion to vanish at long times. This is exactly what we see from the trajectories in Fig.~\ref{fig:trajectory}~(b).

\begin{figure}
	\centering
	\includegraphics[width = 0.45\textwidth]{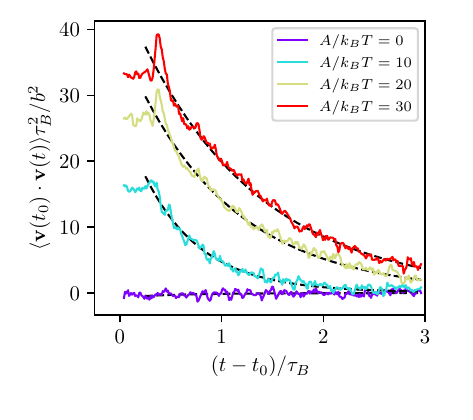}
	\caption{Velocity autocorrelation as a function of time difference $(t-t_0)/\tau_B$ for different values of $A/k_B T$ and $\tau = 0.35 \tau_B$. The initial time $t_0$ has been chosen substantially larger than the time related to the onset of delayed feedback.
	Black dashed lines: exponential fits [see Sec.~\ref{sec:velocity}].\label{fig:velocity_auto}}
\end{figure}

To further illustrate the above-mentioned persistence, we now consider the noise-averaged correlations of the particle's velocity $\mathbf{v}$ at {\em different} times (note that the value at equal times is not accessible due to the divergence of the noise correlation function at zero time difference). For details of the averaging procedure, see Appendix~\ref{sec:averaging}. 
Some results for the correlation function
$\langle \mathbf{v}(t_0)\cdot\mathbf{v}(t)\rangle$ at $\tau=0.35\tau_B$ are shown in Fig.~\ref{fig:velocity_auto}. 
For very small values of $A/k_B T$, i.e. vanishing delay force, the correlation function is essentially zero at all times. This is expected since we hereby approach the limit of a free Brownian particle.
Increasing $A/k_B T$, however, the correlation function assumes large positive values and decays to zero the later the larger $A/k_B T$ is.
We also observe in the regime of short time differences $t-t_0$ the emergence of peaks at multiples of the delay time.
To quantify the decay at large time differences (i.e., after the initial peaks) we have fitted the functions $\langle \mathbf{v}(t_0)\cdot\mathbf{v}(t)\rangle$ 
according to the expression $f(t-t_0) = C_1\exp(-(t-t_0)/\tau_r) + C_2$ where $\tau_r$ is a decay time, and $C_1$ and $C_2$ are constants (note that non-zero values of $C_2$ occur due to the finite time differences considered in Fig.~\ref{fig:velocity_auto}. 
For even larger $t-t_0$ one would expect all functions to approach zero). 
To obtain the exponential fit of the velocity autocorrelation function (see Fig.~\ref{fig:velocity_auto}) we discarded all data points that correspond to times smaller than $t-t_0 = \tau - 0.1\tau_B$ i.e., times before the first visible peak. Physically, it is the history dependence of the feedback which causes short-time oscillations at multiples of the delay time. Only after the first visible peak we see an exponential decrease in the correlation function that justifies the exponential fit.
It is seen from Fig.~\ref{fig:velocity_auto} that this exponential fit describes the data indeed quite well. This means that we can extract (approximate) values for the {\em persistence times} of the Brownian particle under time-delayed feedback. Numerical results from this procedure are given in Table~\ref{tab:persistence}. As an overall trend, we observe an increase of the persistence time with the strength of repulsion. Further, at very large values of $A/k_BT$, $\tau_r$ seems to saturate. We also see that in most cases (apart from $A/k_B T = 5$) the persistence time is larger than the delay time. Similar trends are seen when using a different fitting scheme that we introduce in Sec.~\ref{sec:diffusion}.
\begin{table}
    \centering
    \begin{tabular}{|c|c|c|c|c|c|c|c|c|}
        \hline
        \rule[-1ex]{0pt}{2.5ex} $A/k_B T$ & 5 & 10 & 15 & 20 & 25 & 30 & 35 & 40 \\
        \hline
        \rule[-1ex]{0pt}{2.5ex} $\tau_r/\tau_B$ & 0.25 & 0.56 & 0.81 & 1.00 & 1.08 & 1.19 & 1.35 & 1.34 \\
        \hline
    \end{tabular}
    \caption{Persistence time $\tau_r$ obtained by fitting the velocity autocorrelation function by an exponential expression for different values of $A/k_B T$ and $\tau = 0.35\tau_B$.\label{tab:persistence}}
\end{table}
\subsection{\label{sec:diffusion}Diffusion and effective propulsion}
\begin{figure}
	\centering
	\includegraphics[width = 0.45\textwidth]{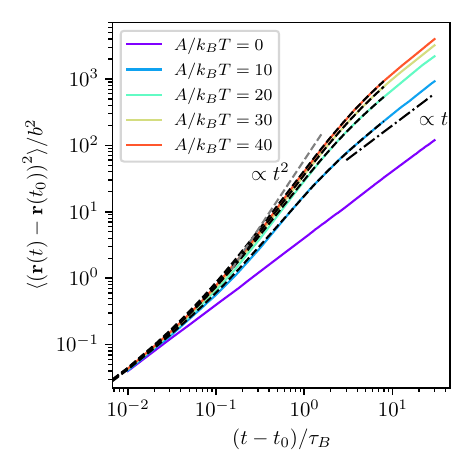}
	\caption{Double-logarithmic plot of the mean-squared displacement (MSD) as a function of time difference $t-t_0$ for different values of $A/k_B T$ and, thus, $A\tau/\gamma b^2$ with $\tau = 0.35 \tau_B$. Dash-dotted line: Serves as a guide for the eye to indicate diffusive behavior ($\propto t$) in the long time limit. Dashed grey line serves as a guide for the eye and represents ballistic behavior, where the MSD is proportional to $t^2$.
		Dashed black lines represent the fit according to the analytical result for the MSD of an active Brownian particle [see Eq. (\ref{eq:ABP_MSD})].\label{fig:msd}}		
\end{figure}

\begin{figure}
	\centering
	\includegraphics[width = 0.45\textwidth]{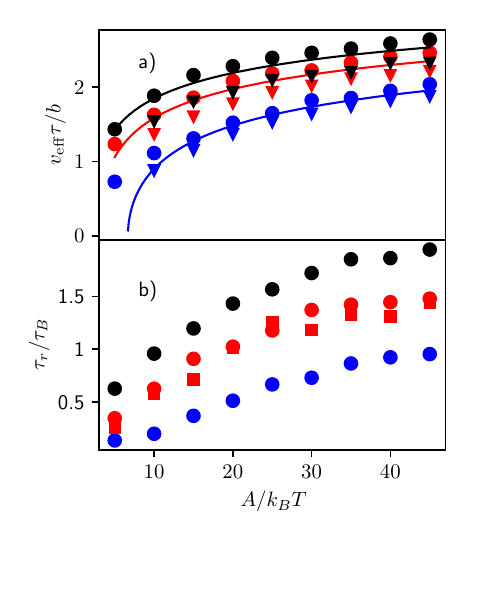}
	\caption{a) Effective velocity as a function of the feedback strength $A/k_BT$ for different values of the delay time (blue: $\tau = 0.15\tau_B$, red: $\tau = 0.35\tau_B$, black: $\tau = 0.55\tau_B$). 
		Circles: Obtained from a fit of the mean-squared displacement (MSD) according to the analytical expression for the active Brownian particle (ABP) model [see Eq. (\ref{eq:ABP_MSD})]. Triangles: Obtained from the mixed regime of the MSD. Solid lines: Long-time velocity of the corresponding deterministic model [see Eq. (\ref{eq:const_vel})].  b) Circles: Persistence time $\tau_r$, obtained from a fit of the MSD according to the ABP model (color code as in a)). Squares: Persistence time $\tau_r$, obtained from an exponential fit of the velocity autocorrelation function [see Sec.~\ref{sec:velocity}] for $\tau = 0.35 \tau_B$.\label{fig:v_eff_tau_r_delay_new}}
\end{figure}

To further characterize the stochastic motion we now consider the mean-squared displacement (MSD),
$\left\langle\left(\mathbf{r}(t)-\mathbf{r}(t_0)\right)^2\right\rangle$. Numerical results for the MSD at various values of feedback strength and fixed delay time $\tau=0.35\tau_B$ are plotted in Fig.~\ref{fig:msd}.
One can identify different regimes. After an initial transient period, all curves exhibit a range of time differences where the MSD exhibits features of both, ballistic motion, i.e., with a quadratic dependence on $t-t_0$, and diffusive motion (with a linear dependence on $t-t_0$).
Given that the stochastic system is overdamped, we take the mixed behavior seen in the MSD as an indication of an effective {\em propulsion} mechanism driving the particle. 
We note in passing that for all parameter combinations considered in Fig.~\ref{fig:msd}, the particle would indeed approach a finite long-time velocity in the deterministic case (since $A\tau/\gamma b^2>1$). Turning back to the stochastic case we find that,
at larger time differences, the MSD crosses over towards {\em diffusive} behavior, i.e. linear dependency on $t$. This is expected from the finite range of the velocity autocorrelation function (Fig.~\ref{fig:velocity_auto}), that is, the randomization of the direction of motion at long times. Assuming that, in the mixed regime, the MSD is proportional to $4Dt + v_\mathrm{eff}^2 t^2$, with
$v_\mathrm{eff}$ being the effective velocity of propulsion, we can extract this parameter directly from the MSD (see Appendix~\ref{sec:velfrommsd} for details). Results from this procedure are shown in Fig.~\ref{fig:v_eff_tau_r_delay_new} a) (triangles). As expected, the effective propulsion velocity increases with the strength of delayed feedback. We also note that the extracted values are somewhat smaller, but still quite close to the long-time velocity in the noise-free case.

The other data (circles) for $v_\mathrm{eff}$ shown in Fig.~\ref{fig:v_eff_tau_r_delay_new} a) have been extracted in a different way. The underlying idea is that
the behavior of the MSD of the particle under repulsive feedback, particularly the intermediate ballistic regime and the subsequent crossover to diffusive motion, is indeed reminiscent of MSD data for active (self-propelled) particles. A prominent theoretical model
	is the so-called active Brownian particle (ABP). Here we aim to establish a link to this particular model, since its MSD is known analytically \cite{howse_self-motile_2007-1, ten_hagen_brownian_2011-1} and its overall physical behavior has been studied extensively \cite{lowen_inertial_2020}. The ABP equations of motion are given by \cite{lowen_inertial_2020,zottl_emergent_2016-1}
\begin{eqnarray}
\gamma\frac{d\mathbf{r}}{dt} &= &\gamma v_0\mathbf{n}(t)+\bm{\xi}(t),\nonumber\\
\dot{\theta}(t)&=&\xi_r(t).
\label{eq:ABP}
\end{eqnarray}
Here, $v_0$ is the speed of self propulsion and $\mathbf{n}(t)=(\cos\theta,\sin\theta)$ is the (unit) heading vector, whose orientational dynamics described by the polar angle $\theta$ is governed by the second member of Eq.~(\ref{eq:ABP}). The translational noise $\bm{\xi}(t)$ is defined as in Eq.~(\ref{eq:delaylangevin}), and the rotational noise (with zero average) fulfills $\langle\xi_r(t)\xi_r(t')\rangle=2D_r\delta(t-t')$, with $D_r$ being the rotational diffusion constant. The latter defines the persistence time of the ABP, $\tau_r=1/D_r$.
The MSD of the ABP can be calculated analytically \cite{howse_self-motile_2007-1, ten_hagen_brownian_2011-1}. The result can be written as \cite{zottl_emergent_2016-1}
\begin{eqnarray}
	\langle\left(\mathbf{r}(t) - \mathbf{r}(t_0)\right)^2\rangle &=& 4D(t-t_0) + 2v_0^2 \tau_r^2 \left(\frac{(t-t_0)}{\tau_r} \right.\nonumber\\
	& &\left.+ \exp\left(-\frac{(t-t_0)}{\tau_r}\right) -1\right).
	\label{eq:ABP_MSD}
\end{eqnarray}
At short times $t-t_0 \ll \tau_r$ the right hand side of Eq.~(\ref{eq:ABP_MSD}) reduces to $4D\left(t-t_0\right) + v_0^2\left(t-t_0\right)^2$, reflecting the ballistic contribution $\propto v_0^2$. In contrast,
for $t-t_0 \gg \tau_r$, the MSD of the ABP becomes purely diffusive, that is, $\langle\left(\mathbf{r}(t)-\mathbf{r}(t_0)\right)^2\rangle = 4D_\mathrm{eff}\left(t-t_0\right)$, with $D_{\mathrm{eff}}=D + v_0^2\tau_r/2$ being the effective diffusion coefficient. 

The idea is now to use Eq.~(\ref{eq:ABP_MSD}) as a fit formula for the MSD of the particle under delayed feedback at hand, with $v_0=v_\mathrm{eff}$ and $\tau_r$ being the fit parameters. To fit the MSD, we discarded data points corresponding to $t - t_0 < 0.07 \tau_B$, and afterwards use all data points up to $t - t_0 = 0.35\tau_B$. Subsequently, every 50th data point up to $t = 8\tau_B$ is used. Generally, we find that the agreement between our MSD and the ABP expression is satisfactory at intermediate and large time differences $t-t_0$, while at small time differences visible deviations appear. The resulting values
for the propulsion velocity are shown in Fig.~\ref{fig:v_eff_tau_r_delay_new} a) (circles). They follow the same trend (i.e., increase with the feedback strength) as the previously discussed velocities extracted directly from the ballistic regime of the MSD.
In fact, the two procedures give quite similar values, with the exact long-time velocities of the deterministic case lying in between. 
The persistence times $\tau_r$ resulting from the ABP fit are shown in Fig.~\ref{fig:v_eff_tau_r_delay_new} b) (circles), together with the data extracted from the velocity autocorrelation function (squares), see Table~\ref{tab:persistence}.
The general behavior seen from the two sets of data is consistent: $\tau_r$ increases with the $A/k_BT$ and tends to saturate for large values of the repulsion strengths. Quantitatively, the values obtained at fixed $A/k_B T$ somewhat differ. One should note, however that the precise values of $\tau_r$ do depend on details of the averaging and fitting procedure (e.g., the start of the fit). 
Disregarding these technical details, we note that the overall trends observed so far for $v_{\mathrm{eff}}$ and $\tau_r$ persist for other delay times. Exemplary data are shown in Fig.~\ref{fig:v_eff_tau_r_delay_new}. An exception is the case $A/k_B T = 5$ for $\tau = 0.15\tau_B$, where the deterministic system would not display a long time velocity at all.

\begin{figure}
	\centering
	\includegraphics[width = 0.45\textwidth]{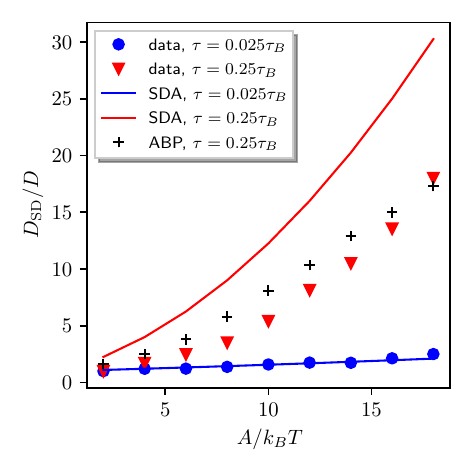}
	\caption{Renormalized diffusion coefficient $D_\mathrm{eff}/D$ extracted from the long-time mean-squared displacement as function of $A/k_B T$ for $\tau = 0.25\tau_B$ (triangles) and $\tau = 0.025\tau_B$ (circles). Included are the predictions from the long time limit of the active Brownian particle relation [see Eq. (\ref{eq:ABP_MSD})] for $\tau = 0.25\tau_B$ (crosses), and the results from the small delay approximation (solid lines) [see Eq.~(\ref{eq:diffusion_sd})].\label{fig:diffusion}}
\end{figure}

To complete the picture, we finally consider the long-time diffusion coefficient governing the motion of the particle under delayed feedback. We extract this coefficient from the MSD data by fitting the linear behavior observed at large time differences $t-t_0$. Numerical data for the renormalized diffusion coefficient $D_\mathrm{eff}/D$ (where $D$ pertains to free Brownian motion)
for different strengths of feedback and two values of the delay time are shown in Fig.~\ref{fig:diffusion}.
As expected, the diffusion coefficient is enhanced relative to the free case for all parameter combinations considered. Furthermore, we observe an increase of $D_\mathrm{eff}/D$ with both, $A/k_B T$ and $\tau$. 
This is expected in view of the corresponding behavior of $v_\mathrm{eff}$, see Fig.~\ref{fig:v_eff_tau_r_delay_new}~a). 
In fact, given the connection to the ABP model utilized earlier, it is interesting to directly compare the present numerical data with the {\em exact} prediction for ABPs \cite{howse_self-motile_2007-1}, $D_{\mathrm{ABP}}=D + v_0^2\tau_r/4$. 
Replacing $v_0$ and $\tau_r$ by the values extracted for $v_\mathrm{eff}$ and $\tau_r$ (by fitting the MSD according to the ABP model) thus gives us another estimate for the long-time diffusion coefficient.
Results for the delay time $\tau=0.25 \tau_B$ and different values of $A/k_B T$ are indicated by crosses in Fig.~\ref{fig:diffusion}. The values are indeed quite close to those obtained by a direct fit of the long-time MSD. 

Finally, given all the numerical results $D_\mathrm{eff}/D$ obtained either directly or by fitting procedures, one may ask about an {\em analytical} estimate, at least for small delay times $\tau$.
To this end we employ the so-called small-delay approximation \cite{frank_stationary_2001-2,guillouzic_small_1999-1,kuchler_langevins_1992-1}). Specifically, to investigate the dynamics up to linear order in $\tau$, we expand the deterministic force 
given in Eq. \eqref{eq:fbforce} up to first order in $\tau$. Noting that the zero-th order term vanishes, we obtain
\begin{equation}
\gamma \dot{\mathbf{r}}(t) \approx -\tau\mathbf{H}(\mathbf{0})\dot{\mathbf{r}}(t) + \bm{\xi}(t),
\end{equation}
where $\mathbf{H}(\mathbf{0})$ is again the (diagonal) Hessian matrix evaluated at zero displacement (with $H_{\alpha\alpha}(0)=-A/b^2$).
Upon reinserting the approximate expression for $\dot{\mathbf{r}}$ and linearizing in $\tau$ we find
\begin{equation}
\gamma \dot{\mathbf{r}}(t) \approx\left(1 + \frac{A}{\gamma b^2}\tau\right)\bm{\xi}(t).
\label{eq:GRfb}
\end{equation}
The right hand side of Eq.~(\ref{eq:GRfb}) has the same structure as the equation of motion for a free Brownian particle, yet with an renormalized diffusion coefficient $D_{\mathrm{SD}}$. To see this more explicitly, we replace
the white noise $\bm{\xi}(t)$ by a noise with unit variance, i.e. $\bm{\eta}(t)=\sqrt{2\gamma k_B T}^{-1}\bm{\xi}(t)$ (with $\langle \eta_\alpha (t) \eta_\beta(t^\prime) \rangle = \delta(t-t^\prime)\delta_{\alpha\beta}$), yielding
\begin{eqnarray}
\dot{\mathbf{r}}(t)& \approx & \gamma^{-1}\left(1 + \frac{A}{\gamma b^2}\tau\right)\bm{\xi}(t)\nonumber\\
&\stackrel{!}{=}& \sqrt{2D_\mathrm{SD}}\bm{\eta}(t).
\label{eq:GRfb2}
\end{eqnarray}
Recalling the definition of the diffusion coefficient of free motion, $D=k_BT/\gamma$, we obtain for the renormalized diffusion coefficient (within the small delay approximation)
	\begin{equation}
		\frac{D_\mathrm{SD}}{D} = \left[1 + \frac{A}{\gamma b^2}\tau\right]^2. 
		\label{eq:diffusion_sd}
	\end{equation}
In Fig.~\ref{fig:diffusion}, the prediction given in Eq.~(\ref{eq:diffusion_sd}) is shown by solid lines. It is seen that the small-delay approximation gives indeed a very good, quantitative description of the real data if the delay time is substantially smaller than the Brownian time ($\tau=0.025\tau_B$).  It also provides the correct trend regarding the role of the delay time, that is, for fixed $A/k_B T$, $D_\mathrm{SD}$ increases with $\tau$.
For the larger delay time $\tau=0.25\tau_B$, however, one observes profound quantitative differences, specifically, an overestimation of the long-time diffusion coefficient. Expectedly, these deviations worsen when $\tau$ increases further.
Still, one may conclude that the small-delay approximation preserves the overall trends seen in the numerical data.

\section{\label{sec:conclusions}Conclusions}
In this paper we have investigated the dynamical behavior of a Brownian particle moving in two dimensions under feedback control with time delay. Specifically, we have chosen a nonlinear, repulsive feedback force with Pyragas-like dependence on the delay time, $\tau$, such that the feedback control vanishes at $\tau=0$. As a consequence, all differences observed relative to the free Brownian case are induced by the delay itself, combined with the strength (and range) of repulsion.

We have seen that the resulting behavior can be indeed quite intriguing and different from the frequently considered case of a {\em linear} feedback force. Already in the deterministic limit, the 
particle can develop a stationary state characterized by a {\em constant} velocity vector with non-zero magnitude and direction. Using a stability analysis we have shown that this state, which is absent in the linear case, is marginally stable. Upon introducing noise, the direction 
of motion randomizes at long times. However, in marked contrast to a free Brownian particle, the feedback-controlled particle shows persistent motion over a finite range of time. Altogether, the resulting behavior in the noisy case (and at appropriate values of delay time and repulsion strength)
resembles that of a self-propelled particle. Here we specifically compared our numerically obtained MSD with that of an active, self-propelled Brownian particle (ABP), which is widely used in theoretical calculations \cite{solon_active_2015} but also in experiments \cite{buttinoni_active_2012-1,bechinger_active_2016}. 
However, in principle, one could have also compared with other models of active particles, such as run-and-tumble \cite{solon_active_2015,cates_when_2013} or active Ornstein-Uhlenbeck \cite{fodor_how_2016}, which show a similar behavior of the MSD. The comparison with the analytical expression for the MSD of the ABP model allowed us to extract values for the persistence time $\tau_r$ and the propulsion velocity $v_\mathrm{eff}$ as functions of delay time and repulsion strength. It turns out that $v_\mathrm{eff}$ is close to the analytical result of the deterministic model.

To conclude, the present model for time-delayed repulsive feedback can be interpreted as a mechanism of propulsion with feedback-dependent persistence.
In this sense, our work contributes to establishing a link between systems under time-delayed feedback  and active matter. Such connections are indeed expected due to fundamental reasons since both types of systems involve non-reciprocal couplings \cite{loos_irreversibility_2020} (in the ABP, the position is affected by the direction and not vice versa. In the feedback-controlled case, the non-reciprocal coupling occurs between the hidden variables describing the delayed feedback) and are typically out of equilibrium \cite{loos_irreversibility_2020}. 
For that reason an analysis of the thermodynamic properties of the present model would be very interesting \cite{loos_heat_2019,loos_irreversibility_2020}.

Our work can also be seen as an example of a soft matter system where the delay is not an annoyance, but rather works in a constructive manner: indeed, the persistent motion only occurs for finite time delay. Similar constructive effects of time delayed feedback \cite{holubec_finite-size_2021-1,tarama_traveling_2019,loos_heat_2019,bielawski_controlling_1994} have recently been identified in studies of active Janus particles \cite{liebchen_clustering_2015,muinos-landin_reinforcement_2021}.

Clearly, it would be very interesting to compare our predictions to experiments. As stated in the introduction, the present model is not directly aimed at a specific experimental set-up. Still, we expect that the proposed nudging effects could be realized, e.g., with optical forces. 

We also note of a seemingly different system where such an effect occurs: in a recent study of cell migration on viscoelastic substrates \cite{clark_self-generated_2022}, the retarded creation of bump-like perturbations was identified as a source of directed motion that can be described with a model quite similar to ours.

Finally, given the similarities of our model to active motion on the single-particle level, it seems very promising to use the model in a many-particle set-up. Indeed, in view of the wealth of studies on the collective behavior of ABPs revealing, e.g., motiliy-induced clustering \cite{buttinoni_dynamical_2013,bialke_active_2015} and local alignment of motion \cite{caprini_spontaneous_2020}, it would be exciting to see whether the present model shows similar behavior. Work in this direction is under way.
Another route would be to apply the present control scheme to an active (rather than a passive) Brownian particle. In fact, the study of active particles with retardation effects due to feedback, inertia or other mechanisms is an emerging field, where first steps have already been done \cite{mijalkov_engineering_2016,muinos-landin_reinforcement_2021,holubec_finite-size_2021-1,sprenger_active_2022,wang_spontaneous_2022}.

\begin{acknowledgments}
We gratefully acknowledge the support of the Deutsche Forschungsgemeinschaft (DFG, German Research Foundation), project number 163436311 - SFB 910.
Furthermore we would like to thank Bernold Fiedler for inspiring discussions concerning the stability analysis of the deterministic system. 
\end{acknowledgments}

\appendix
\section{\label{sec:technical}Numerical details}
Simulations were done using the C++ programming language. Data analysis was done using the python programming language and the python package {\em matplotlib} \cite{Hunter:2007} for visualization.
\subsection{\label{sec:bd}Brownian Dynamics simulations}
To solve the stochastic SDDE (\ref{eq:sdde}) numerically we performed BD simulations based on the Euler-Maruyama integration scheme \cite{maruyama_continuous_1955,kloeden_numerical_1992}.
The discretized equation of motion reads
\begin{widetext}
    \begin{align}
        \mathbf{r}_{n+1} &= \mathbf{r}_{n} + \gamma^{-1}\mathbf{F}(\mathbf{r}_{n}, \mathbf{r}_{n-N_\tau})\Delta t + \gamma^{-1}\sqrt{2k_B T \gamma \Delta t}~\bm{\eta}_{n+1}\\
        &= \mathbf{r}_{n} + \gamma^{-1}\frac{A}{b^2}(\mathbf{r}_{n} - \mathbf{r}_{n-N_\tau})\exp\left(\frac{-(\mathbf{r}_{n} - \mathbf{r}_{n-N_\tau})^2}{2b^2}\right)\Delta t + \gamma^{-1}\sqrt{2k_B T \gamma \Delta t}~\bm{\eta}_{n+1},
    \end{align}
\end{widetext}
where $N_\tau$ is the number of time steps within one delay time $\tau$, $\Delta t$ is the size of the
time step, and $\bm{\eta}$ is a vector of random numbers drawn from uncorrelated standard normal distributions.
Note that an initial history has to be imposed for the time interval $[-\tau, 0]$ via the discretized history function $\mathbf{\Phi}(t_n) = \mathbf{\Phi}_{t_n}$, i.e. the trajectory on the discretized time interval $[-\tau, 0]$ has to be known. 
For all stochastic simulations in this study we used trajectories of free Brownian particles. 
Further, we used $\Delta t = 10^{-5}\tau_B$, which is sufficiently small to yield reasonable numerical precision in all cases considered.
We note, however, that this choice requires significant storage. To give an example, for the delay time $\tau = 0.35 \tau_B$, positional data of $35000$ time steps have to be stored in the memory. 

\subsection{\label{sec:averaging}Averaging}
Here we describe the averaging procedures used to compute the MSD and the velocity autocorrelation function discussed in Secs.~\ref{sec:diffusion} and \ref{sec:velocity}, respectively.
To start with, the MSD was computed according to the equation
\begin{equation}
\langle\left(\mathbf{r}(t) - \mathbf{r}(t_0)\right)^2\rangle = \frac{1}{N}\sum_{k=1}^{N} \left(\mathbf{r}(t) - \mathbf{r}(t_0)\right)^2,
\end{equation}
where $N$ is the number of noise realizations. For the MSD, $N = 8000$ has proven to be sufficient. Further, $t_0$ is the (discrete) starting time.
This time is chosen such that the system has already evolved under time delayed feedback for a rather long time (specifically, $t = 10\tau_B-2\tau$) such that transient behavior has essentially disappeared.
For the velocity autocorrelation the averaging problem is slightly more involved due to the noisy character of the underlying velocity vectors (obtained by numerical differentiation of the positional data).
Here we used a two-step procedure. The first step involves a moving average after computing the velocity by numerical differentiation. Specifically, for every realization and every time step we average over the current and the three following time steps, i.e.,
\begin{equation}
    \mathbf{v}_\mathrm{av}(t) = \frac{1}{4}\sum_{n=0}^{n_\mathrm{max}} \mathbf{v}(t+n\Delta t).
\end{equation}
We set $n_\mathrm{max}= 3$.
We have tested that the actual value of $n_\mathrm{max}$ used in the moving average does not crucially influence the results. As a second step, we perform a noise average (with $N = 80000$, i.e., a factor of ten larger than for the MSD) to obtain the velocity correlation 
function 
\begin{equation}
    \langle\mathbf{v}(t_0)\cdot\mathbf{v}(t)\rangle = \frac{1}{N}\sum_{k=1}^{N} \mathbf{v}_{\mathrm{av},k}(t_0)\cdot\mathbf{v}_{\mathrm{av},k}(t),
\end{equation}
where $t>t_0$.
The calculations are started after the system has evolved under time delayed feedback for a rather long time (specifically, $t = 20\tau_B-2\tau$), even longer than for the computation of the MSD, to be confident that transient behavior has essentially disappeared.
\subsection{\label{sec:velfrommsd}Extraction of effective velocity from MSD}
The average velocity magnitude is obtained from the MSD by considering its mixed regime. 
We assume that 
\begin{equation}
	\mathrm{MSD}(t_2) - \mathrm{MSD}(t_1) = v_\mathrm{eff}^2 (t_2 - t_1)^2 - 4D(t_2 - t_1), 
\end{equation}
where $\mathrm{MSD}(t_1) = \langle\left(\mathbf{r}(t_1) - \mathbf{r}(t_0)\right)^2\rangle$, $\mathrm{MSD}(t_2) = \langle\left(\mathbf{r}(t_2) - \mathbf{r}(t_0)\right)^2\rangle$ and $D$ is the diffusion coefficient of a free Brownian particle. 
We then solve this equation for the effective velocity magnitude $v_\mathrm{eff}$ and obtain 
\begin{equation}
	v_\mathrm{eff} = \sqrt{\frac{\langle\left(\mathbf{r}(t_2) - \mathbf{r}(t_0)\right)^2\rangle - \langle\left(\mathbf{r}(t_1) - \mathbf{r}(t_0)\right)^2\rangle - 4D(t_2-t_1)}{(t_2 - t_1)^2}},
\end{equation}
where $t_2 > t_1$, and both $t_1$ and $t_2$ are within the mixed regime of the MSD.
\section{\label{sec:appendix_solution}Explicit solution of the linear, deterministic case}
As mentioned in Sec.~\ref{sec:linear}, the linear delay equation (\ref{eq:linear}) can be solved by the method of steps (e.g. \cite{driver_ordinary_1977,erneux_applied_2009-1}), with the general solution for the particle's position in first time interval $t\in[0,\tau]$ given in Eq.~(\ref{eq:solution_linear}).
Here we provide the explicit solution for a history function $\mathbf{\Phi}(t)$, $t\in[-\tau,0]$ with linear time dependence. Specifically, the components are given by $\Phi_\alpha=v_0t$ with $v_0=const$, $v_0>0$. Note that as argued before in Sec.~\ref{sec:linear} we can safely work in one dimension in the linear case. The position as function of time up to $t=2\tau$ then reads
	\begin{widetext}
		\begin{equation}
			x(t) = \begin{cases}
				v_0 t, & \hspace{5pt} -\tau < t \leq 0 \\
				v_0\left[\ainv{} + (t - \tau)  + \left(\tau - \ainv{}\right)\exp\left(\anorm{t}\right)\right], & \hspace{5pt} 0 < t \leq \tau \\
				v_0 \left[\ainv{2} + (t-2\tau) + \left( \tau-\ainv{}\right)\exp\left(\anorm{t}\right) + \left(t + \anorm{\tau^2} - \anorm{t\tau} - \ainv{}\right)\exp\left(\anorm{\left(t-\tau\right)}\right)\right], &  \hspace{5pt} \tau < t \leq 2\tau
			\end{cases}
		\end{equation}
	\end{widetext}
%
%
\section{\label{sec:perturbations}Impact of perturbations in the deterministic, nonlinear case}
\begin{figure}
	\centering
	\includegraphics[width = 0.45\textwidth]{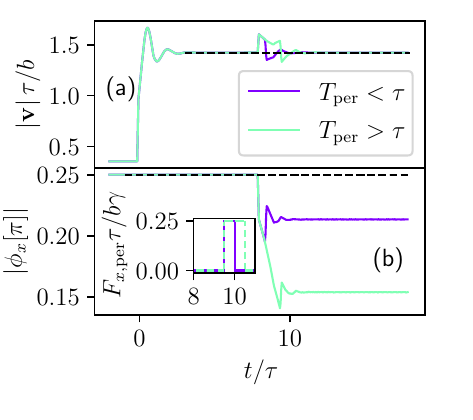}
	\caption{Nonlinear model, deterministic case in 2D: Absolute value of the velocity $|\bm{v}|$ (a) and of the angle $\phi$ between the velocity vector and the $x$-axis (b) for two perturbations (see inset) $F_\mathrm{x,per}$ with different duration $T_\mathrm{per}$ along the $x$-axis ($A\tau/\gamma b^2 = 2.75$).\label{fig:perturbation}}
\end{figure}

As shown by the (linear) stability analysis in Sec.~\ref{sec:longtime}, the steady state characterized by a non-zero velocity is marginally stable. To better understand this issue we present
in Fig.~\ref{fig:perturbation} numerical data illustrating the impact of a time-dependent perturbation. The perturbation acts in a finite time interval starting when the system is already in the stationary state. Specifically, we consider a situation where the vector $\mathbf{v}_\infty$ is characterized by magnitude
$|\mathbf{v}_\infty| = \frac{\sqrt{2}b}{\tau}\sqrt{-\ln(\frac{\gamma b^2}{A\tau})}$ (where $A\tau/\gamma b^2 = 2.75$), and an angle of $\pi/4$ with respect to the positive $x$-axis.
To perturb the system we use a  force $\mathbf{F}_\mathrm{per}(t)$ with rectangular time dependence. Specifically,
\begin{equation}
    \mathbf{F}_\mathrm{per}(t)=
    \begin{cases}
        (0.25,0)^T \frac{\tau}{b\gamma} & \text{if } 9.5\tau < t < 9.5 \tau+T_\mathrm{per},\\
        (0,0)^T & \text{else},
    \end{cases}
\end{equation}
where $T_\mathrm{per}$ is the duration of perturbation. Here we set $T_\mathrm{per}=0.5\tau$ or $1\tau$.
The equation of motion then reads
\begin{equation}
    \gamma \frac{d\mathbf{r}}{dt} = \mathbf{F}(\mathbf{r}(t), \mathbf{r}(t - \tau)) + \mathbf{F}_\mathrm{per}(t).
\end{equation}
As seen from Fig.~\ref{fig:perturbation}~(a), the velocity magnitude is only shortly affected by the perturbation: after some oscillations, it quickly returns to its stationary value. The same behavior is seen for larger magnitudes of the perturbing force. This occurs for both durations of the perturbation. Similar behavior is seen for other shapes of the perturbation.
In contrast, the direction of motion (Fig.~\ref{fig:perturbation}~(b)) is {\em permanently} changed, and the degree of this change depends on the duration of the perturbation. To summarize,
the perturbation changes the characteristics of the long-time state (that is, the angle), but not the fact that a stationary state is reached at all.
\section{\label{sec:veff}Dependence of effective velocity on the delay time}
\begin{figure}
	\centering
	\includegraphics[width = 0.45\textwidth]{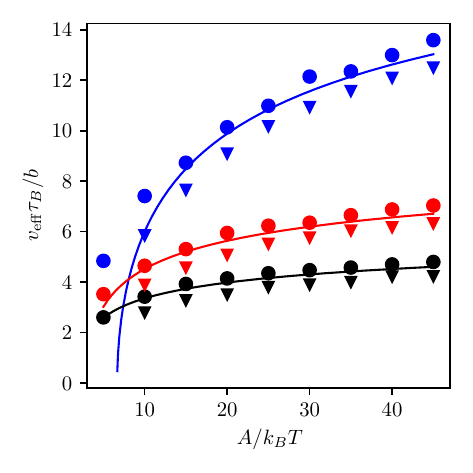}
	\caption{Effective velocity as a function of the feedback strength $A/k_BT$ for different values of the delay time (blue: $\tau = 0.15\tau_B$, red: $\tau = 0.35\tau_B$, black: $\tau = 0.55\tau_B$) in units of $b/\tau_B$. 
			Circles: Obtained from a fit of the mean-squared displacement (MSD) according to the active Brownian particle model [see Eq.~\ref{eq:ABP_MSD}]. Triangles: Obtained from the mixed regime of the MSD [see Sec.~\ref{sec:diffusion}]. Solid lines: Long time velocity of the corresponding deterministic model [see Eq.~(\ref{eq:const_vel})].\label{fig:v_eff_tau_B}}
\end{figure}

		When studying the effective velocity $v_\mathrm{eff}$ as function of the feedback strength, the time scale used to non-dimensionalize this quantity is of crucial importance. Specifically, in Fig.~\ref{fig:v_eff_tau_r_delay_new} we have shown that when plotting $v_\mathrm{eff}$ normalized as $v_\mathrm{eff}\tau/b$, an increase of $\tau$ leads to an increase of $v_\mathrm{eff}\tau/b$ for all values of $A/k_B T$. In contrast, as shown in Fig. \ref{fig:v_eff_tau_B}, the dimensionless velocity $v_\mathrm{eff}\tau_B/b$ shows the opposite trend, that is, it is largest for the smallest delay time at all values of $A/k_B T$.

\bibliography{apssamp}
\end{document}